\documentclass[a4paper,aps,pre,floatfix,twocolumn,nofootinbib,showpacs,superscriptaddress]{revtex4-1}

\usepackage{times}
\usepackage{bbold}
\usepackage{bbm}
\usepackage[normalem]{ulem}
\usepackage[english]{babel}
\usepackage{microtype}
\usepackage[pdftex]{graphicx}
\setkeys{Gin}{width=7.5cm}
\usepackage{latexsym,amsmath}
\usepackage[pdftex]{color}
\usepackage[hidelinks,unicode=true]{hyperref}
\hypersetup{colorlinks=true,linkcolor=blue,urlcolor=blue,citecolor=blue,pdfhighlight=/N}

\usepackage{float}
\usepackage[all]{xy}
\usepackage{comment}
\usepackage{lipsum}

\newcommand{\av}[1]{\langle #1 \rangle}

\begin{document}

\title{Scalar model of flocking dynamics on complex social networks}

\author{M. Carmen Miguel}

\affiliation{Departament de F\'{\i}sica de la Mat\`{e}ria Condensada,
  Universitat de Barcelona, Mart\'{\i} i Franqu\`es 1, 08028 Barcelona,
  Spain}

\affiliation{Universitat de Barcelona Institute of Complex Systems
  (UBICS), Universitat de Barcelona, Barcelona, Spain}

\author{Romualdo Pastor-Satorras}

\affiliation{Departament de F\'isica, Universitat Polit\`ecnica de
  Catalunya, Campus Nord B4, 08034 Barcelona, Spain}

\date{\today}

\begin{abstract}
  We investigate the effects of long-range social interactions in flocking
  dynamics by studying the dynamics of a scalar model of collective motion
  embedded in a complex network representing a pattern of social interactions,
  as observed in several social species. In this scalar model we find a
  phenomenology analogous to that observed in the classic Vicsek model: In
  networks with low heterogeneity, a phase transition separates an ordered from
  a disordered phase. At high levels of heterogeneity, instead, the transition
  is suppressed and the system is always ordered. This observation is backed up
  analytically by the solution of a modified scalar model within an
  heterogeneous mean-field approximation. Our work extends the understanding of
  the effects of social interactions in flocking dynamics and opens the path to
  the analytical study of more complex topologies of social ties.
\end{abstract}

\maketitle

\section{Introduction}
\label{sec:introduction}

The collective motion of interacting mobile agents can lead to stunning
self-organized spatio-temporal patterns, such as those observed in flocks of
birds, shoals of fish or herds of mammals at the macroscopic scale, or in
colonies of bacteria, migrating cells or self-propelled nano-particles at the
microscopic scale~\cite{Vicsek2012,Ramaswamy2010,Cavagna2018}. The study of
collective motion, and in particular the flocking behavior of animals, started
attracting interest decades ago, the first studies dated in 1987, when
C.~W. Reynolds simulated the flocking of birds in terms of an artificial life
simulation using \textit{boids}~\cite{Reynolds:1987:FHS:37402.37406}. The
interest for this phenomenology in the statistical physics community ignited
with the flocking model introduced by Vicsek and coworkers~\cite{Vicsek1995} in
1995. In the so-called Vicsek model, self-propelling particles (SPPs) move in a
$d=2$ space with constant speed and interact among them by aligning their
velocity to the average velocity of a set of other SPPs in their close
neighborhood. The addition of a source of noise $\eta$, accounting for the
physical difficulties in gathering and/or processing local information, leads to
a dynamic phase transition separating an ordered phase at $\eta \leq \eta_c$, in
which particles move coherently in the same average direction, from a disordered
phase at $\eta > \eta_c$, in which SPPs move essentially as uncorrelated
persistent random walkers~\cite{mendez14}. The Vicsek model has allowed to draw
useful analogies between the collective motion of
animals~\cite{Sumpter2006,Sumpter10}, and the well-known features of
order-disorder phase transitions in classical statistical
mechanics~\cite{Vicsek2012,Ginelli2016}.

The analysis of the Vicsek and other flocking models has been performed mostly
in Euclidean spaces~\cite{Ginelli2016}, where the neighborhood of a SPP is
defined in a metric way, given by all other SPPs within a distance $R$ centered
at the original SPP. Other works have also considered nonmetric neighborhoods,
given by the SPPs in the first shell in a Voronoi tessellation constructed from
the position of the particles at each time step~\cite{Ginelli2010}. This sort of
interactions assume that all individuals (SPPs) are equivalent, and therefore
simplify numerical and analytical approximations~\cite{Toner1995}. This
simplification, however, comes at the cost of disregarding the possible effects
of social interactions, that can cause individuals to follow preferentially
those others with which they have strong social ties~\cite{Ling2019}. The
effects of such social ties, represented in terms of non-metric pairwise
interactions encoded in a complex network~\cite{Newman10}, have been considered
in the framework of the Vicsek model in different
works~\cite{Bode2011,Bode2011a,SEKUNDA2016,Miguel2017}. In particular,
Ref.~\cite{Miguel2017} studied the effects of a heterogeneous complex topology,
as represented by a scale-free degree distribution (defined as the probability
that a node is connected to $k$ others, or has degree $k$) with a power-law
form~\cite{Barabasi:1999}, $P(k) \sim k^{-\gamma}$, on the order-disorder
transition experimented by the model. In this work, it was reported that for
$\gamma > 5/2$, a standard transition is observed for a finite value of $\eta$,
while for $\gamma < 5/2$, the transition is suppressed in the thermodynamic
limit of infinite network size, being the system in the ordered state for all
physical values of the noise parameter $\eta$\footnote{In the standard Vicsek
  model, the range of physical values of $\eta$ is restricted to a finite
  interval that can be taken to be $0 \leq \eta \leq 1$~\cite{Vicsek1995}. In
  Ref.~\cite{Miguel2017} it was shown that $\eta_c \to 1$ in the
  thermodynamic limit for $\gamma < 5/2$.}. This result is relevant for the
understanding of the flocking behavior of a variety of social animals whose
social contact networks~\cite{croft2008exploring} have been reported to have
scale-free signatures~\cite{Lusseau2003,Manno2008}, and it indicates that
flocking is more robust against external fluctuations in the case of high
network heterogeneity (i.e.~small $\gamma$).

The results presented in~\cite{Miguel2017} were backed up by numerical
simulations and argued to be related with the behavior observed analytically in
the majority-vote (MV) model in
networks~\cite{deOliveira1992,PhysRevE.71.016123,Huepe2002,Aldana2004}. In the
MV model, nodes are endowed with binary spin variables, taking values
$\{ +1, -1\}$. With probability $1-f$, nodes copy the spin orientation of the
majority of their neighbors, while with probability $f$ they adopt the orientation
opposite to that of the majority.  A dynamic phase transition is observed in the
MV at a noise threshold $f_c$, separating an ordered phase for $f < f_c$ from a
disordered one at $f > f_c$.  Analytical calculations on scale-free networks,
based on a heterogeneous mean-field (HMF)
theory~\cite{pv01a,dorogovtsev07:_critic_phenom}, provide the expression of the
threshold~\cite{Chen2015}
\begin{equation}
  \label{eq:19}
  f_c = \frac{1}{2} - \frac{1}{2} \sqrt{\frac{\pi}{2}}
  \frac{\av{k}}{\av{k^{3/2}}},
\end{equation}
where $\av{k^n} = \sum_k k^n P(k)$ is the $n$-th moment of the degree
distribution. This formula implies that for $\gamma >5/2$ a threshold
$f_c < 1/2$ is obtained, while for $\gamma<5/2$, the noise threshold takes its
maximum value $f_c = 1/2$. This value, corresponding to a completely disordered
system, indicates that the system is always ordered in the thermodynamic limit.
The behavior of the MV model is thus akin to that observed in the Vicsek model
on networks, and one can argue that they are equivalent in the sense that the
dimensionality of the order parameter appears to be irrelevant in the
characterization of the behavior of critical transitions in
networks~\cite{dorogovtsev07:_critic_phenom,Miguel2017}.

In this paper we delve further into the role of a complex topology on collective
motion by considering the model of flocking dynamics proposed by Czir\'ok,
Barab\'asi and Vicsek (CBV)~\cite{Czirok1999}. In the CBV model particles move
in a one-dimensional ring with a velocity represented by a real number,
$u_i \in \mathbb{R}$, which tends to align with the average velocity of other
particles in a local neighborhood, and which is affected by a random noise of
amplitude $\eta$. This model, characterized by a continuous scalar order
parameter, has been used to model the marching behavior of swarms of
locusts~\cite{Buhl2006,Ariel2015}. Numerical simulations as well as analytical
calculations based on a continuous hydrodynamic description, show that the CBV
model in one dimension experiences a dynamic phase transition, separating an
ordered phase at low noise from a disordered one at high noise. The observed
non-equilibrium phase transition is characterized by a set of exponents
different from those observed in the vectorial standard Vicsek model in
$d=2$~\cite{Czirok1999}.  Here we show, by means of extensive numerical
simulations, that a complex topology affects the CBV model in a way analogous to
the Vicsek model: In scale-free topologies with degree exponent $\gamma > 5/2$,
a standard transition is preserved; on the other hand, for $\gamma < 5/2$, the
transition is absent in the thermodynamic limit. The two models, however, show
differences in the critical exponents computed at the transition point. Our
numerical analysis is complemented by the analytical solution within the HMF
approximation of a modification of the CBV model, showing that the critical
noise in the thermodynamic limit is proportional to the moment $\av{k^{3/2}}$ of
the degree distribution, and thus drastically changing its behavior when
$\gamma$ crosses the value $5/2$. The HMF approximation allows to compute the
critical exponent $\beta$ in the ordered phase, which is found to be different
to the one obtained for the MV model.  Our results extend the range of analyses
of collective motion models on networks, and suggest that the presence of the
crossover degree exponent $\gamma = 5/2$ might be a general feature of flocking
models on networks based on averaging rules applied to nearest neighbors.

\section{The CBV model in networks}\label{sec:barab-vics-flock}

The CBV model~\cite{Czirok1999} was originally defined on a $d=1$ space with
periodic boundary conditions, in which particles can move with velocity $u_i$.
Each particle $i$ updates its velocity by taking the average $\av{u}_i$ of the
velocity of other particles in a neighborhood
$\mathcal{V}_i = [x_i - \Delta, x_i + \Delta]$ surrounding it, i.e.
$\av{u}_i = \sum_{j | x_j \in \mathcal{V}_i} u_j / N_i$, where $N_i$ in the
number of particles in $\mathcal{V}_i$. The average velocity is modulated by a
function $G(u)$, such that, when $u>1$, $G(u) < u$, and when $u<1$, $G(u) > u$,
and symmetrically for negative velocities, in such a way to force the modulus of
the velocities to stay close to $1$.  The function used in
Ref.~\cite{Czirok1999} is
\begin{eqnarray}
  \label{eq:6}
   G(u) = \frac{u + \mathrm{sign(u)}}{2},
\end{eqnarray}
where $\mathrm{sign(u)}$ is the sign function.
Finally, a noise term $\eta \xi_i$ is added, where $\eta$ gauges the noise
strength and $\xi_i$ is a random number uniformly distributed in the interval
$[-1/2, 1/2]$.

In the case of interactions mediated by a network, the topology is defined by
the adjacency matrix $a_{ij}$, taking value $1$ if nodes $i$ and $j$ are
connected by an edge, and zero otherwise~\cite{Newman10}.  We define the
dynamical update rule for velocities as
\begin{equation}
  u_i(t+1) = G \left[ \frac{\sum_j a_{ij} u_j(t)}{k_i} \right] + \eta
  \xi_i.
  \label{eq:13}
\end{equation}
With this prescription, we do not consider the interaction of a node with
itself.  This particular choice does not have a strong effect on the dynamics of
the model, with the exception of possibly reducing the critical value of the
noise. Obviously, the prescription presents problems for nodes of degree $1$,
which simply consider their only neighbor.  To avoid this problem, in the
following we will consider a sufficiently large minimum degree. As in the case
of the vectorial Vicsek model on networks, the position of particles plays no
role and we do not keep track of them~\cite{Miguel2017}.

The order parameter of the model, $\phi(\eta)$ is defined in terms of the time
average of the  average velocity
\begin{equation}
  \phi_\eta(t) = \frac{1}{N}  \sum_i u_i(t),
\end{equation}
namely
\begin{equation}
  \phi(\eta) = \av{\phi_\eta(t) }_t   \equiv \lim_{T\to\infty}\frac{1}{T} \int_{0}^{T}
  \phi_\eta(\tau) \; d \tau.
\end{equation}
In this model we expect the presence of a critical point $\eta_c$ separating a
disordered phase with $\phi(\eta) = 0$ for $\eta > \eta_c$ from an ordered or
flocking one at $\eta < \eta_c$, in which
$\phi(\eta) \sim (\eta_c - \eta)^\beta$, defining the critical exponent $\beta$.

\section{Numerical analysis}
\label{sec:numerical-analysis}

In order to check the behavior of the CVB model, we have performed extensive
numerical simulations on uncorrelated power-law networks with degree
distribution $P(k) \sim k^{-\gamma}$ and minimum degree $m = 3$, generated using
the uncorrelated configuration model (UCM)~\cite{Catanzaro05}. To estimate the
order parameter we perform averages over $250000$ time steps, after letting the
dynamics thermalize for a sufficiently large time. Due to the velocity reversal
invariance of the dynamics, we actually compute the order parameter as the time
average of the absolute value of average velocity,
$\phi(\eta) = \av{| \phi_\eta(t) |}_t$.

\begin{figure}[t]
  \includegraphics{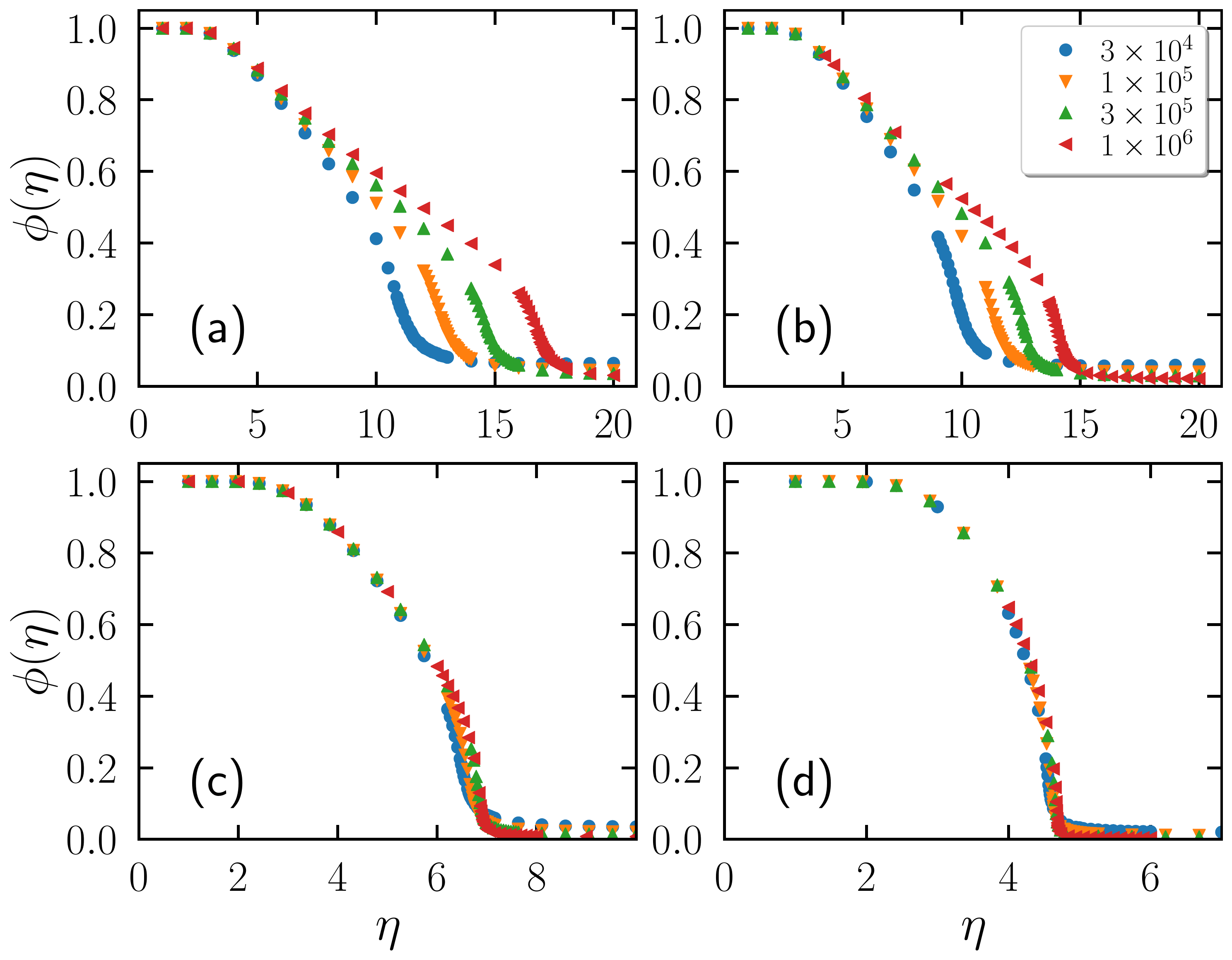}
  \caption{Order parameter $\phi(\eta)$ as a function of $\eta$ in the CBV model
    on UCM networks of different size $N$. Panels correspond to different values
    of the degree exponent: (a) $\gamma = 2.10$, (b) $\gamma=2.20$, (c)
    $\gamma = 2.75$, (d) $\gamma=3.50$.}
  \label{fig:orderparameter}
\end{figure}

In Fig.~\ref{fig:orderparameter} we show the order parameter $\phi(\eta)$ as a
function of the noise intensity $\eta$ computed in networks of different degree
exponent $\gamma$ and size $N$. As we can see from this Figure, in the region
$\gamma > 5/2$ the curves $\phi(\eta)$ seem to collapse for a sufficiently large
network size, while for $\gamma < 5/2$ the value of $\eta$ at which the order
parameter becomes zero increases with $N$. This behavior is fully compatible
with the observations made for the vectorial Vicsek model. One difference stands
out, however. In the definition of the Vicsek model, the maximum physical value
of the noise is $\eta=1$. In the CBV model, instead, $\eta$ is unbounded, so we
expect the critical noise $\eta_c$ to diverge with network size for
$\gamma < 5/2$.

\begin{figure}[t]
  \includegraphics{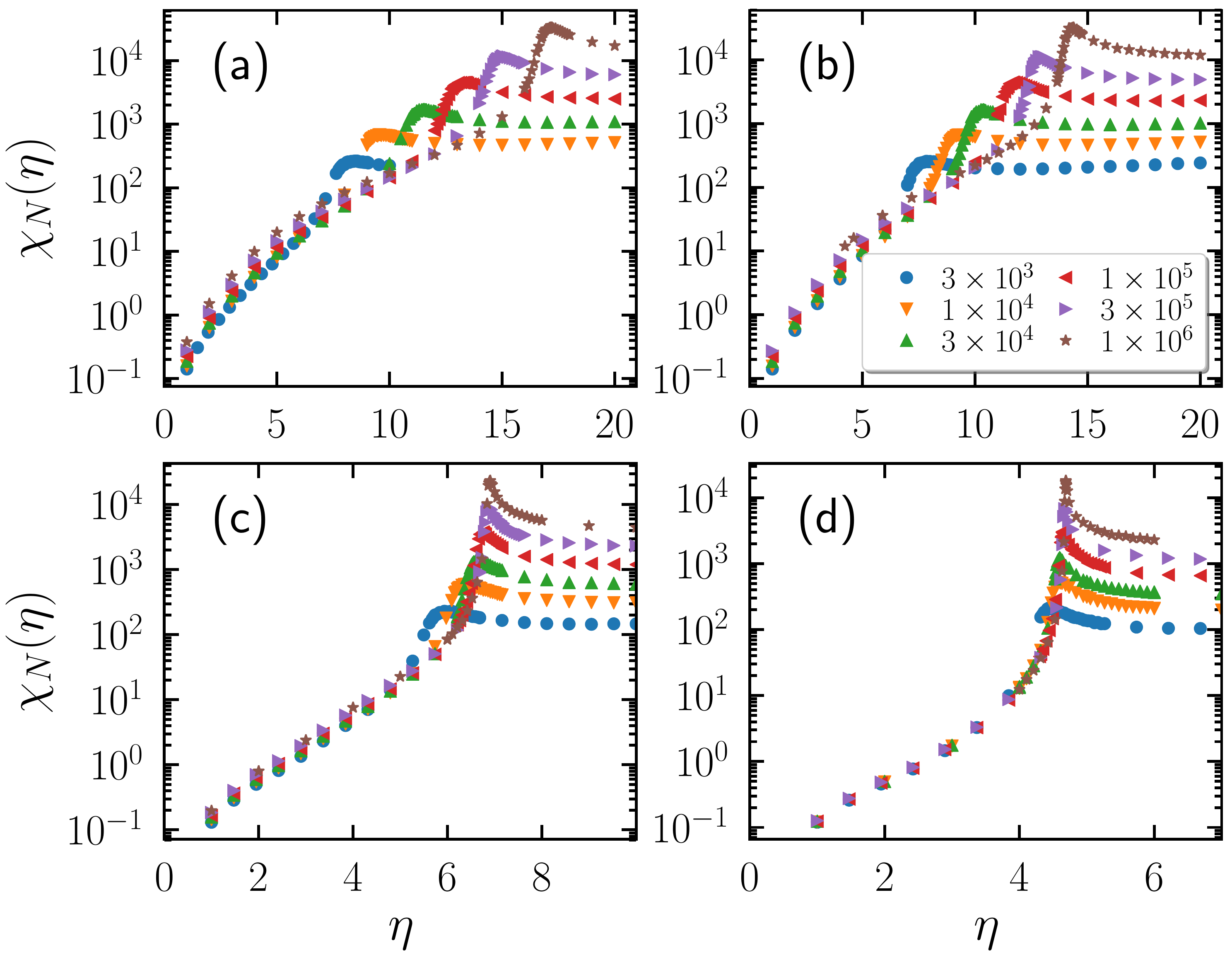}
  \caption{Dynamic susceptibility $\chi_N(\eta)$ as a function of $\eta$ in the
    CBV model on UCM networks of different size $N$. Panels correspond to
    different values of the degree exponent: (a) $\gamma = 2.10$, (b)
    $\gamma=2.20$, (c) $\gamma = 2.75$, (d) $\gamma=3.50$.}
  \label{fig:susceptibility}
\end{figure}

In order to further check this expectation, we have computed the dynamic
susceptibility~\cite{Ferreira2012,Castellano2016,Miguel2017}
\begin{equation}
  \label{eq:4}
  \chi_N(\eta) = N \frac{\av{\phi_\eta(t)^2}_t -
    \phi(\eta)^2}{\phi(\eta)}.
\end{equation}
In the presence of a second order phase transition, the dynamic susceptibility
develops a peak at a value $\eta_c(N)$ that is interpreted as the size
dependent critical point in a network of fixed size $N$. In
Fig.~\ref{fig:susceptibility} we plot the dynamic susceptibility computed from
different $\gamma$ and $N$ values.  For all values of $\gamma$ considered, the
dynamic susceptibility shows the presence of a sharp peak, signaling the
corresponding finite-size critical point that for $\gamma >5/2$ tends to a
constant value, while it diverges for $\gamma<5/2$.

\begin{figure}[t]
  \includegraphics{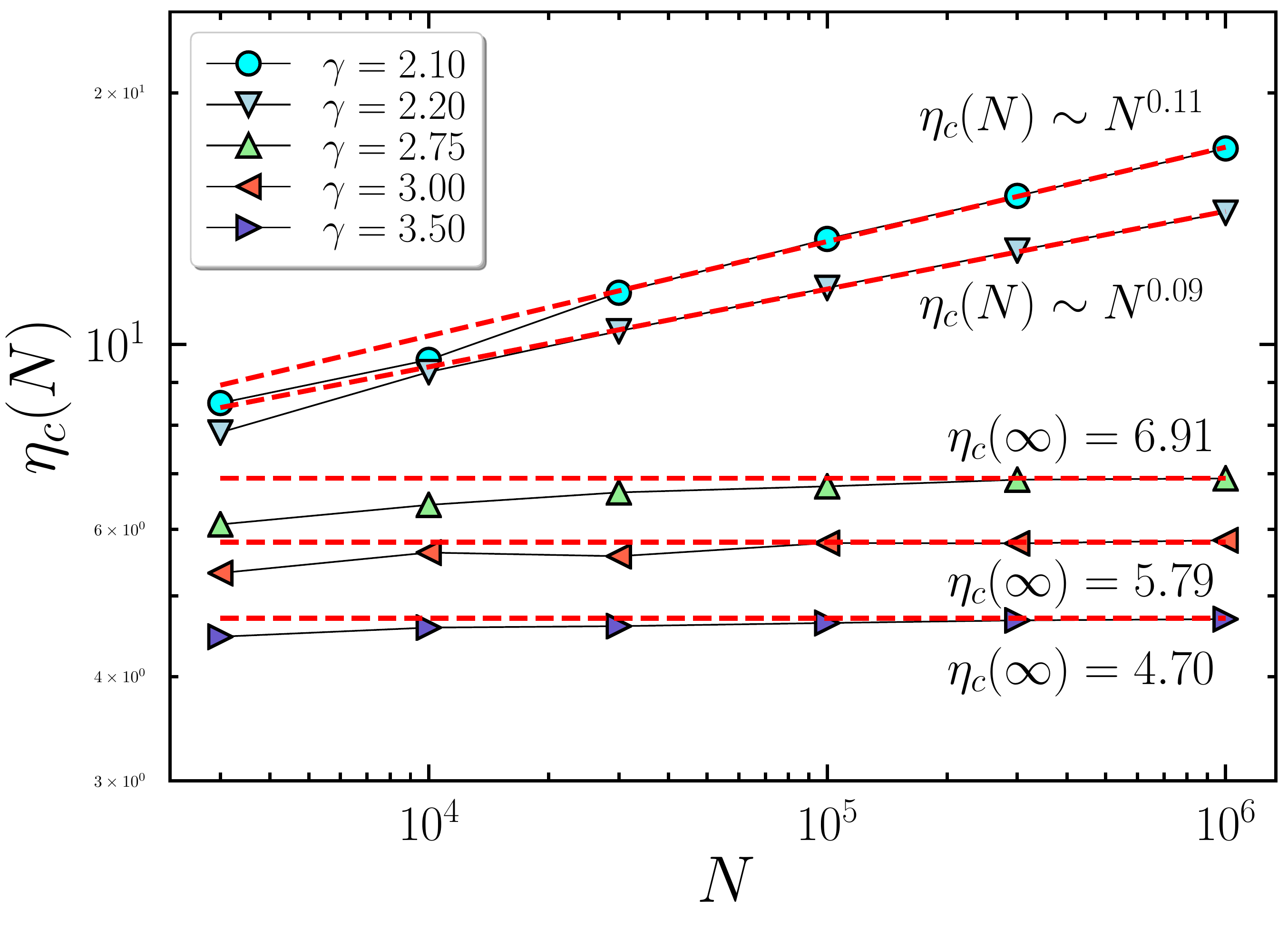}
  \caption{Peak of the dynamic susceptibility $\eta_c(N)$ as a function of
    network size $N$ in the CBV model on UCM networks of different degree
    exponent $\gamma$. Dashed lines represent the critical point at the
    thermodynamic limit estimated using a finite-size scaling ansatz
    ($\gamma > 5/2$) and a power-law fit to the last four points
    ($\gamma < 5/2$).}
  \label{fig:peak_vs_N}
\end{figure}

In Fig.~\ref{fig:peak_vs_N} we plot the value of the effective critical point
$\eta_c(N)$ estimated from the peak of the susceptibility as a function of the
network size $N$.
For $\gamma > 5/2$, the peaks tend to a constant value for increasing $N$. We
plot as dashed lines the critical point in the thermodynamic limit,
$\eta_c(\infty)$, estimated by applying a finite size scaling of the form
$\eta_c(N) = \eta_c(\infty) - a N^{-b}$~\cite{cardy88}.  The values estimated by
performing a non-linear regression of the numerical data to this ansatz form are
$\eta_c(\infty) = 6.91(1)$ for $\gamma=2.75$, $\eta_c(\infty) = 5.79(1)$ for
$\gamma=3.00$, and $\eta_c(\infty) = 4.70(1)$ for $\gamma=3.50$, which are
compatible with the peaks obtained for the largest size $N=10^6$ considered. For
$\gamma < 5/2$, on the other hand, the effective thresholds seems to grow as a
power-law with $N$. A linear regression to the form $\eta_c(N) \sim N^{b}$,
performed over the four largest network sizes, leads to the exponent $b=0.11$
for $\gamma=2.10$ and $b=0.09$ for $\gamma=2.20$. The effective critical point
$\eta_c(N)$ growing as a power-law of the network size indicates an infinite
critical noise in the thermodynamic limit, fully compatible with the maximal
noise $\eta_c = 1$ in the thermodynamic limit obtained in the Vicsek model.

\begin{figure}[t]
  \includegraphics{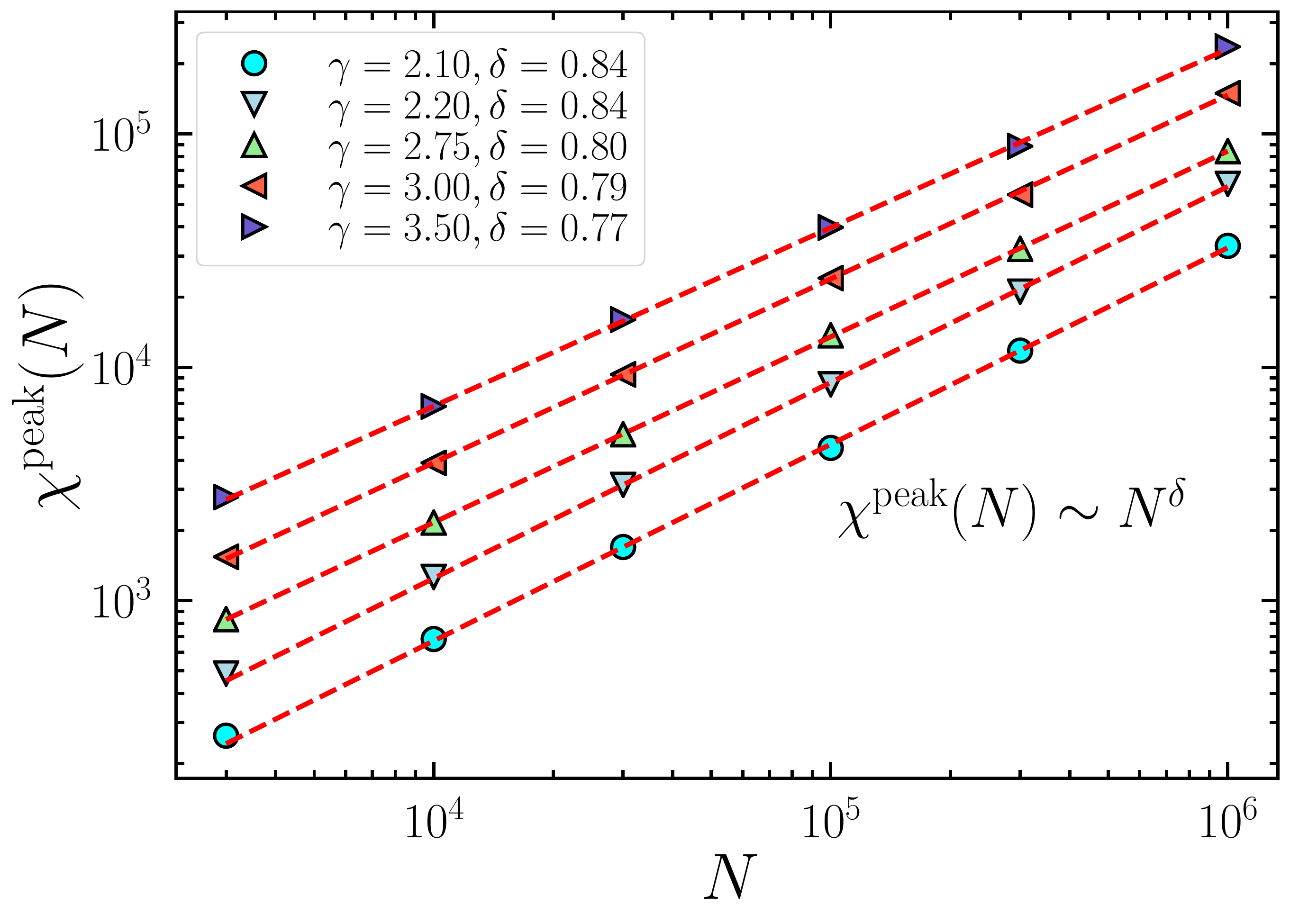}
  \caption{Scaling of the peak of the dynamic susceptibility
    $\chi^\mathrm{peak}(N)$ as a function of network size $N$ in the CBV model
    on UCM networks of different degree exponent $\gamma$.  The exponents
    $\delta$ quoted are obtained by means of a linear regression to the form
    $\chi^\mathrm{peak}(N) \sim N^{\delta}$. Plots have been shifted vertically
    for the sake of clarity.}
  \label{fig:height_vs_N}
\end{figure}

Another signature of criticality in the order-disorder transition of the CBV
model on networks is given by the height of the peak of the dynamic
susceptibility $\chi^\mathrm{peak}(N)$, which is expected to scale as a
power-law of the network size,
$\chi^\mathrm{peak}(N) \sim N^\delta$~\cite{Ferreira2012,Castellano2016}, where
$\delta$ is another characteristic critical exponent. In Ref.~\cite{Miguel2017}
it was observed numerically that the exponent $\delta$ for the vectorial Vicsek
model on scale-free networks takes, as a function of $\gamma$, the same values
as the MV model. This observation was used to strengthen the relation between
both models. In Fig.~\ref{fig:height_vs_N} we present the results of the scaling
analysis of $\chi^\mathrm{peak}(N)$ for the CVB model.
For $\gamma < 5/2$, the exponent $\delta$ seems to be constant, and
approximately equal to $0.84(2)$. For $\gamma > 5/2$, the exponents are smaller,
taking the value $\delta = 0.80(2)$ for $\gamma=2.75$ $\delta = 0.79(2)$ for
$\gamma=3.00$, and $\delta=0.77(1)$ for $\gamma=3.50$. This last result is
compatible with the mean-field value obtained for the MV model,
$\delta_\mathrm{MF} = 3/4$~\cite{PhysRevE.71.016123}. The exponents for
$5/2 < \gamma < 3$ appear constant and approximately equal to $0.80$, but this
observation could be a finite size effect.

Interestingly however, and despite the coincidence for large $\gamma$ values, we
observe that the exponents obtained for the CBV at small $\gamma$ are different
from those arising in the Vicsek model. In particular, from
Ref.~\cite{Miguel2017}, we have for the Vicsek model $\delta = 0.57$ for
$\gamma=2.10$, clearly incompatible with the value obtained here. This
observation prompts the conclusion that, while the Vicsek and CBV model show an
analogous scaling of the threshold as a function of $\gamma$, they do not share
the same critical exponents in scale-free networks.

\section{Heterogeneous mean-field analysis}\label{sec:heter-mean-field}

Even though the model defined in the previous Section is simpler in essence to
the original Vicsek model in networks, it still cannot be solved
analytically. To be able to get some insight, we perform a further
simplification based upon the observation in~\cite{Czirok1999} that the actual
shape of the function $G(u)$ is irrelevant, not even its lack of continuity at
$u=0$. Inspired by this observation, we consider here a modified CBV model in
which the modulating function $G(u)$ is a simple majority rule
\begin{equation}
  G(u) = \mathrm{sign(u)} = \left\{
    \begin{array}{cc}
      +1 & \mathrm{if} \; u > 0\\
      -1 & \mathrm{if} \; u <0
    \end{array}
  \right. .
\end{equation}
With this definition, the dynamics amounts to computing the average velocity of
the nearest neighbors, keeping its sign, and adding a noise term.

We can recast the dynamics of this modified CBV model in terms of a set of
\textit{dual} variables $u_i^*(t)$ as
\begin{eqnarray}
  \label{eq:16}
  u_i^*(t+1) &=& G[\{ u_i(t)\}], \label{eq:5}\\
  u_i(t+1) &=&  u_i^*(t+1) + \eta \xi_i \label{eq:18},
\end{eqnarray}
where $\{ u_i(t)\}$ denotes the set of possible values of $u_i$, $i \in [1, N]$.
From this prescription, it is easy to see by direct substitution of
Eq.~\eqref{eq:18} into Eq.~\eqref{eq:5} that the dual variables $u_i^*$ fulfill
the dynamic update
\begin{equation}
  u_i^*(t+1) = G[\{ u_i^*(t) +  \eta \xi_i\}].
  \label{backward}
\end{equation}
The difference of this prescription with respect to Eq.~(\ref{eq:13}) is that,
in the original case, the final velocities are a real number, while in the
modified expression the dual velocities $u_i^*(t)$ are binary variables,
restricted to be $\{-1, +1\}$. From the dual variables, the original velocity
values can be obtained as
\begin{equation}
  \label{eq:30}
  u_i(t) = u_i^*(t) + \eta \xi_i.
\end{equation}

Eq.~(\ref{backward}) can be further simplified in the case of a function $G(u)$
given by the sign function. In this case, the normalization factor $k_i^{-1}$ in
Eq.~(\ref{eq:13}) can be omitted, since it does not affect the sign of the
argument, so we have the final dynamical rule
\begin{equation}
  \label{eq:17}
   u_i^*(t+1) = G \left[ \sum_j a_{ij} u_j^*(t) + \eta \sum_j a_{ij} \xi_j
   \right].
\end{equation}
With the new prescription in Eq.~(\ref{eq:17}), an order parameter
$\phi^*(\eta)$ can be computed over the dual velocities $u_i^*(t)$. Since the
noise $\xi_i$ has zero mean, the order parameters $\phi^*(\eta)$ and
$\phi(\eta)$ coincide in the thermodynamic limit, and therefore the critical
properties of the model can be computed from any prescription. In the following,
we will consider the dual dynamics in Eq.~\eqref{eq:17}, neglecting the star
superindex to simplify notation.

We can solve the modified CBV model with the prescription given in
Eq.~(\ref{eq:17}) at the level of the dual variables by applying an
heterogeneous mean-field (HMF)
approach~\cite{pv01a,Pastor-Satorras:2014aa,dorogovtsev07:_critic_phenom}
inspired in the resolution of similar spin models in
networks~\cite{Castellano2006,Huepe2002,Aldana2004,Chen2015}.  To proceed, let
us define $\rho_k(t)$ the probability that a randomly chosen node of degree $k$
is in state $+1$ at time $t$, and $\psi_k(t)$ as the probability that a node of
degree $k$ flips to the state $+1$ at time $t$. Notice that, given
Eq.~\eqref{eq:17}, the probability $\psi_k(t)$ is independent of the state of
the node $k$ considered, and depends only on the state of its nearest
neighbors. These two quantities are related by the rate equation
\begin{eqnarray}
  \dot{\rho}_k(t) &=& - \rho_k(t) [1 - \psi_k(t)] + [1 - \rho_k(t)] \psi_k(t) \\
            &=& - \rho_k(t) + \psi_k(t),
\end{eqnarray}
which in the steady state $\dot{\rho}_k(t) = 0$ leads to
\begin{equation}
  \rho_k = \psi_k.
\end{equation}

To compute $\psi_k$, we consider the process defined by the dual variables: A
node $i$ of degree $k$ looks at its $k$ nearest neighbors, and considers their
values $\{-1, +1\}$, plus the addition of a random number $\eta \xi_j$, to
compute the quantity
\begin{equation}
  S_i(k) = \sum_j a_{ij} u_j + \sum_j \eta a_{ij} \xi_j.
  \label{eq:22}
\end{equation}
Adopting the annealed network approximation~\cite{dorogovtsev07:_critic_phenom}
to estimate this quantity in the steady state, we define the probability $Q$
that an edge departing from any node arrives at a node in state $+1$. In
uncorrelated networks~\cite{alexei,assortative}, we have
\begin{equation}
  Q = \frac{1}{\av{k}} \sum_k k P(k) \rho_k.
  \label{eq:10}
\end{equation}
The sum $S_i(k)$ can be decomposed in the contribution of the velocities of the
nearest neighbors, $S_{nn} = \sum_j a_{ij} u_j$, and the noise terms,
$S_\eta = \eta \sum_j a_{ij} \xi_j$. $S_{nn}$ results from the addition of the
values $+1$ or $-1$ of $k$ neighbors. Of this $k$ neighbors, there are $n$ in
state $+1$ with probability
\begin{equation}
  P(n) = \binom{k}{n} Q^n (1-Q)^{k-n}.
\end{equation}
Thus,
\begin{equation}
  S_{nn}(n) = (+1) \times n + (-1) \times (k-n) = 2n-k,
\end{equation}
with probability $P(n)$, for $n=0, 1, \ldots k$.

In turn $S_\eta$ is the result of the addition of $k$ random numbers. Assuming
that the noise values $\xi_j$ are independent random variables, and neglecting
correlations between $S_\eta$ in different nodes, we can apply the central limit
theorem and consider that $S_\eta$ is a Gaussian random variable of zero mean
and variance $k \sigma^2$, where $\sigma^2 = \eta^2 / 12$ is the variance of the
original uniform noise distribution in the interval $[-\eta/2, \eta/2]$.
Therefore, we have that $S_\eta = r \in [-\infty, \infty]$ with probability
\begin{eqnarray}
  \label{eq:1}
  P(r) = \frac{1}{\sqrt{2 \pi k \sigma^2}}e^{-r^2/(2 k \sigma^2)}.
\end{eqnarray}
Applying the dynamical rule Eq.~(\ref{eq:17}), the probability that the
variable $u_i$ is flipped to $+1$ is equal to the probability that the sum
$S_i(k)$ is larger than zero. This happens for $r > k -2n$, that takes place
with probability
\begin{eqnarray}
  \psi_k &=& \sum_{n=0}^k \binom{k}{n}Q^n (1-Q)^{k-n} \int_{k-2n}^\infty
  \frac{1}{\sqrt{2 \pi k \sigma^2}}e^{-r^2/(2 k \sigma^2)}\; dr \nonumber \\
             &=& \sum_{n=0}^k \binom{k}{n}Q^n (1-Q)^{k-n}
                \left[ \frac{1}{2} - \mathrm{erf}\left( \frac{ k -2n}{\sqrt{2 k}
                 \sigma}  \right)\right],
    \label{eq:2}
\end{eqnarray}
where $\mathrm{erf}(z)$ is the error function~\cite{abramovitz}.
Eq.~(\ref{eq:2}) can be simplified using the Gaussian approximation for
the binomial distribution, valid when $k$ is large. Thus, the binomial
with parameters $k$ and $Q$ can be approximated by a normal with mean
$\mu_b = k Q$ and variance $\sigma_b^2 = k Q (1-Q)$. Replacing the summation
by an integral in Eq.~(\ref{eq:2}), we have
\begin{eqnarray}
  \label{eq:3}
  \psi_k &=& \frac{1}{2} - \frac{1}{\sqrt{2 \pi \sigma_b^2}}
  \int_{-\infty}^{\infty} e^{- (\mu_b-n)^2/ (2 \sigma_b^2)}
  \mathrm{erf}\left( \frac{ k -2n}{\sqrt{2 k} \sigma}  \right) \; dn \nonumber
  \\
&=& \frac{1}{2} +  \mathrm{erf}\left(  \frac{k y}{ \sqrt{2}
      \sigma_b \sqrt{1 + \frac{k \sigma^2}{4 \sigma_b^2}}} \right),\label{eq:8}
\end{eqnarray}
where we have defined the new variable $y = 1/2 - Q$ that maps the disordered
state, corresponding to $Q=1/2$, into $y=0$.  Close to the disordered state, the
limit of small $y$ leads to
$\sigma_b = \sqrt{k} (1 - 4 y^2)^{-1/2}/2 \simeq \sqrt{k}/2$. Inserting this
approximation into Eq.~(\ref{eq:8}), we finally have
\begin{equation}
  \label{eq:9}
  \psi_k = \frac{1}{2} +  \mathrm{erf}\left(  \lambda\sqrt{2 k} y
       \right),
\end{equation}
where we have defined
\begin{equation}
  \lambda \equiv  \frac{1}{\sqrt{1+\sigma^2}}
  \label{eq:21}
\end{equation}
for later convenience.

We can use Eq.~\eqref{eq:9} to self-consistently solve for $y$ in
Eq.~(\ref{eq:10}), considering that in the steady state $\rho_k = \psi_k$. Since
$y = Q - 1/2$, we have, from Eq.~(\ref{eq:10}),
\begin{equation}
  \label{eq:11}
  y = \frac{1}{\av{k}} \sum_k k P(k)  \mathrm{erf}\left(  \lambda \sqrt{2 k}
    y  \right) \equiv \Psi(y).
\end{equation}
For small $\lambda$, when disorder dominates, the only solution is $y=0$
($Q=1/2$). For a sufficiently large $\lambda$, instead, a stable symmetric
solution $y \neq 0$ appears. The onset of this solution takes place then
\begin{equation}
  \label{eq:12}
  \left .\frac{d\Psi(y)}{dy} \right |_{y=0} > 1.
\end{equation}
Performing the derivative in Eq.~(\ref{eq:11}), we obtain
\begin{eqnarray}
  \left .\frac{d\Psi(y)}{dy} \right |_{y=0}
  &=&  \frac{1}{\av{k}} \sum_k k P(k)
      \left .\frac{d }{dy} \mathrm{erf}\left(  \lambda\sqrt{2 k}
      y  \right)  \right |_{y=0} \nonumber\\
  &=& \frac{2 \sqrt{2}}{\sqrt{\pi}} \lambda
      \frac{\av{k^{3/2}}}{\av{k}} > 1.
\end{eqnarray}
This relation defines the threshold
\begin{equation}
  \label{eq:20}
  \lambda_c = \frac{1}{2} \sqrt{\frac{\pi}{2}} \frac{\av{k}}{\av{k^{3/2}}},
\end{equation}
such that for $\lambda > \lambda_c$ there is an ordered state in the system. In
terms of the noise intensity $\eta$, the ordered states takes place for
$\eta < \eta_c$, defining the threshold
\begin{equation}
  \label{eq:15}
  \eta_c = \sqrt{12}\left[  \frac{8}{\pi} \left(  \frac{\av{k^{3/2}}}{\av{k}}
    \right)^2 -1  \right]^{1/2}.
\end{equation}
For $\gamma < 5/2$ and large $N$, $\av{k^{3/2}}$ diverges and we have
$\eta_c \sim \av{k^{3/2}} / \av{k}$, a growing function of $N$. This indicates
that, in the thermodynamic limit, there is no transition and the system is
always ordered. On the other hand, for $\gamma > 5/2$, $\av{k^{3/2}}$ is finite
and there is always a transition at some finite value of $\eta_c$.

It is interesting to note that the threshold for the MV model in networks,
Eq.~\eqref{eq:19}, can be written as
\begin{equation}
  \label{eq:47}
  f_c = \frac{1}{2} - \lambda_c,
\end{equation}
reflecting an evident relation between the MV and modified CBV models.  This
relation seems natural, given the equations defining the MV and modified CBV
models. However, we have not been able to find a microscopic mapping between the
two models.

\begin{figure}[t]
  \includegraphics{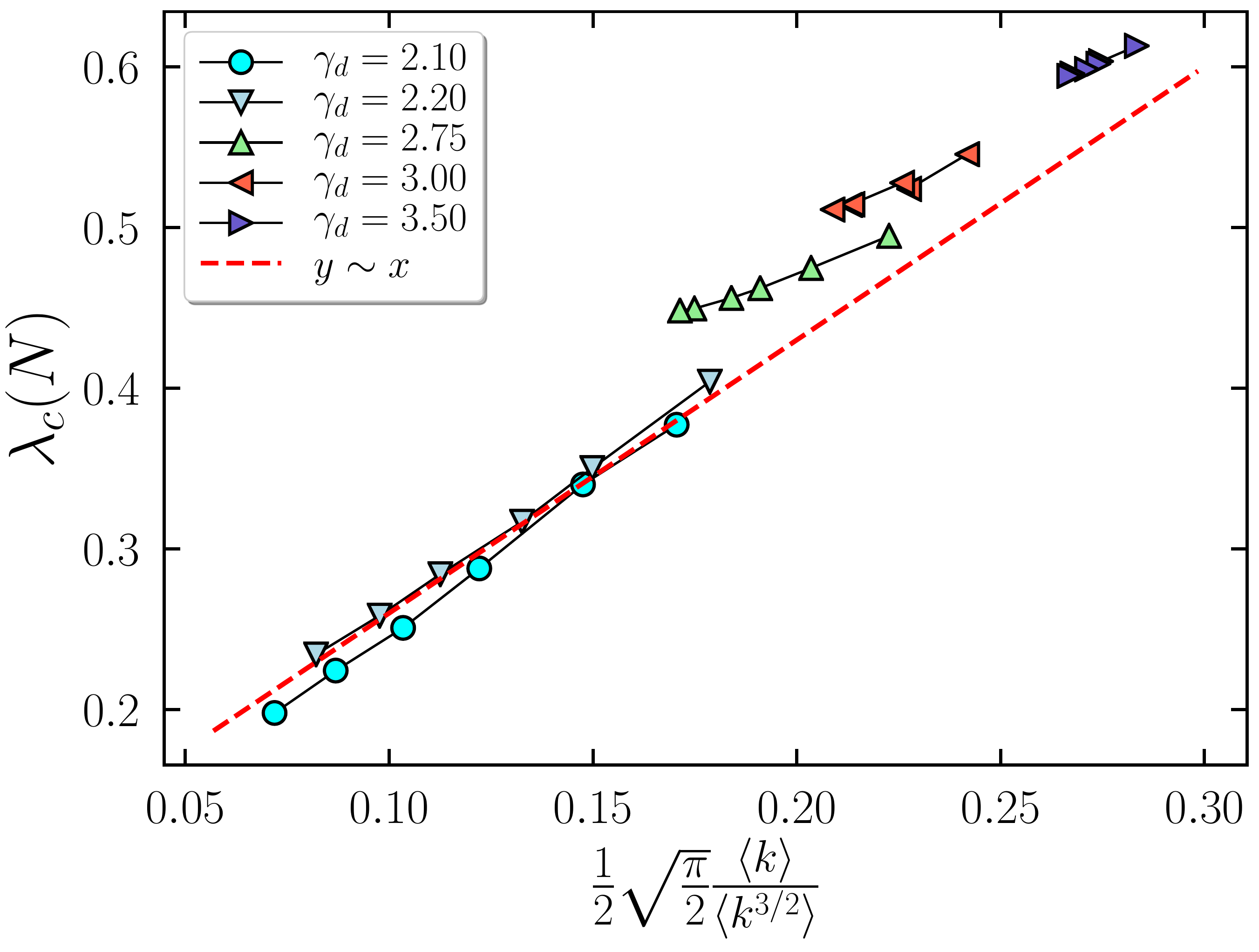}
  \caption{Rescaled peak of the dynamic susceptibility $\lambda_c(N)$ as a
    function of the theoretical prediction, Eq.~\eqref{eq:20}, in the CBV model
    on UCM networks of different degree exponent $\gamma$ and size $N$.}
  \label{fig:peak_vs_theory}
\end{figure}

We can compare the theoretical prediction of the modified CBV model against the
results of computer simulations of the original CBV model presented in
Sec.~\ref{sec:numerical-analysis}. The HMF analysis suggests that the natural
scaling of the CBV dynamics is given in terms of the $\lambda$ parameter,
defined as (see Eq.~\eqref{eq:21})
\begin{equation}
  \label{eq:48}
  \lambda(\eta) = \frac{1}{\sqrt{1+\eta^2/12}}.
\end{equation}
In Fig.~\ref{fig:peak_vs_theory} we present a plot of the parameter $\lambda$ at
criticality, $\lambda_c(N) = (1+\eta_c(N)^2/12)^{-1/2}$, as a function of the
theoretical prediction
$\lambda_c = \frac{1}{2} \sqrt{\frac{\pi}{2}} \frac{\av{k}}{\av{k^{3/2}}}$,
estimated from a direct numerical evaluation of the moments $\av{k}$ and
$\av{k^{3/2}}$ in the networks considered.
As we can see from this figure, while the theoretical prediction for the
modified model provides a very good description of the correct scaling of the
critical point for $\gamma < 5/2$, it is still off regarding the slope by a
factor around $1.7$.  For $\gamma>5/2$ the prediction is not very accurate for
the system sizes considered, but it nevertheless hints towards a linear behavior
for larger $N$.

We can also compute the $\beta$ exponent in the ordered phase of the modified
CBV model within the HMF approximation. The order parameter for the dual
variables, $\phi(\eta) = \frac{1}{N} \sum_{i}u^*_i$, can be written, assuming
the steady state condition $\rho_k = \psi_k$, as
$\phi(\eta) = \sum_k P(k) \left[ 2\psi_k - 1 \right]$.  Using the expression for
$\psi_k$ in Eq.~\eqref{eq:9}, we have
\begin{equation}
  \label{eq:7}
  \phi(\eta) = 2\sum_k P(k) \mathrm{erf}\left(  \lambda\sqrt{2 k} y
       \right).
\end{equation}
Within the continuous degree approximation, replacing summations by integrals,
and assuming $P(k) = (\gamma-1)m^{\gamma-1} k^{-\gamma}$, where the degree $k$
extends from the minimum value $m$ up to infinity, the equations to solve can be
written as
\begin{eqnarray}
  \label{eq:14}
  y &=& I(\gamma-1, y),\\
  \phi(\eta) &=& 2 I(\gamma, y),
\end{eqnarray}
where we have defined
\begin{eqnarray}
  \label{eq:23}
  I(\alpha, y) &\equiv& (\alpha-1)m^{\alpha-1} \int_m^\infty k^{-\alpha}
  \mathrm{erf}\left(  \lambda\sqrt{2 k} y
  \right) dk \\
  &=& \mathrm{erf}\left(z \right) + \frac{z^{2(\alpha-1)}}{\sqrt{\pi}} \Gamma
      \left( \frac{3}{2} - \alpha, z^2 \right),
\end{eqnarray}
where $z = \lambda \sqrt{2 m} y$, and $\Gamma \left( a, z \right)$ is the
incomplete Gamma function~\cite{abramovitz}. Expanding to lowest order the error
and incomplete Gamma functions for small $z$ we have
\begin{eqnarray}
  \label{eq:25}
  \sqrt{\pi} I(\alpha) &=& \frac{4(\alpha-1)}{2\alpha-3} z -
                           \frac{4(\alpha-1)}{3(2\alpha-5)} z^3 \nonumber \\
  &+& \Gamma\left( \frac{3}{2} -\alpha
  \right)z^{2(\alpha-1)}  + \mathcal{O}(z^4),
\end{eqnarray}
where $\Gamma(a)$ is the Gamma function~\cite{abramovitz}.

Since $\gamma>2$, $2(\gamma-1) > 2$, and therefore the lowest order term for the
order parameter is
\begin{equation}
  \label{eq:26}
   \phi(\eta) = 2 I(\gamma, y) \simeq \frac{8(\gamma-1)}{2\gamma-3}
   \frac{z}{\sqrt{\pi}} = \frac{8(\gamma-1)}{2\gamma-3} \lambda
   \sqrt{\frac{2m}{\pi}} y,
\end{equation}
linear in the parameter $y$. To find the value of $y$ we have to solve the
self-consistent equation $y = I(\gamma-1, y)$, which, close to $y=0$, can be
written as
\begin{equation}
  \label{eq:36}
  \sqrt{\pi} y \simeq \frac{4(\gamma-2)}{2\gamma-5} z -
  \frac{4(\gamma-2)}{3(2\gamma-7)}z^3 + \Gamma\left( \frac{5}{2} -\gamma \right)
  z^{2(\gamma-2)}.
\end{equation}
The solution of this equation depends on the value of $\gamma$:
\begin{itemize}
\item $\gamma > \frac{7}{2}$.
  In this range, the leading behavior is
  \begin{equation}
    \label{eq:37}
    \sqrt{\pi} y \simeq \frac{4(\gamma-2)}{2\gamma-5} z -
    \frac{4(\gamma-2)}{3(2\gamma-7)}z^3,
  \end{equation}
  whose non-zero solution is
  \begin{equation}
    \label{eq:38}
    y^2 \simeq \frac{3(2\gamma-7)}{2m\lambda^3(2\gamma-5)} \left( \lambda -
      \lambda_c \right),
  \end{equation}
  where
  \begin{equation}
    \label{eq:39}
    \lambda_c = \sqrt{\frac{\pi}{2m}} \frac{2\gamma-5}{4(\gamma-2)},
  \end{equation}
  in agreement with Eq.~\eqref{eq:20} in the continuous degree approximation. From
  Eqs.~\eqref{eq:38} and~\eqref{eq:26} we have
  \begin{equation}
    \label{eq:40}
    \phi(\eta) \sim (\lambda - \lambda_c)^{1/2}.
  \end{equation}

\item $\frac{5}{2} < \gamma < \frac{7}{2}$.  Here, the leading
  behavior in the equation for $y$ is
  \begin{equation}
    \label{eq:41}
    \sqrt{\pi} y \simeq  \frac{4(\gamma-2)}{2\gamma-5} z - \frac{2 \Gamma\left(
        \frac{7}{2} - \gamma \right)}{2\gamma-5} z^{2(\gamma-2)}.
  \end{equation}

  The non-zero solution is, in this case,
  \begin{equation}
    \label{eq:42}
    \left(\lambda \sqrt{2 m} y\right)^{2\gamma-5} \simeq \frac{4(\gamma-2)}{2
      \lambda
      \Gamma\left(
        \frac{7}{2} - \gamma \right)} \left( \lambda - \lambda_c \right),
  \end{equation}
  with $\lambda_c$ given by Eq.~\eqref{eq:39}. This implies the scaling of the
  order parameter
  \begin{equation}
    \label{eq:43}
    \phi(\eta) \sim (\lambda - \lambda_c)^{1/(2\gamma-5)}.
  \end{equation}

  \item  $\gamma<\frac{5}{2}$. The leading behavior corresponds to the non-zero
    solution
    \begin{equation}
      \label{eq:44}
      y^{5-2\gamma} \simeq \frac{\Gamma\left(
        \frac{5}{2} - \gamma \right)}{(2m)^{2-\gamma}\sqrt{\pi}}
    \lambda^{2{\gamma-2}},
  \end{equation}
  that corresponds to an order parameter
  \begin{equation}
    \label{eq:45}
    \phi(\eta) \sim \lambda^{1/(5-2\gamma)}.
  \end{equation}
  In this case, the critical threshold is $\lambda_c = 0$, in agreement with
  Eq.~\eqref{eq:20} in the infinite network limit, and corresponding to a noise
  threshold $\eta_c \to \infty$.
\end{itemize}

The critical properties of the modified CBV model at the HMF level are
summarized in Table~\ref{tab:critical_values}.  Interestingly, these exponents
differ from those found in Ref.~\cite{Aldana2004} for the MV, namely
$\beta = 1/(2\gamma-3)$ for $3/2 < \gamma < 5/2$ and $\beta = 1/2$ for
$\gamma > 5/2$.

\begin{table}[b]
  \begin{ruledtabular}
    \begin{tabular}{|l|c|c|}
      $\gamma$ &  $\lambda_c$ & $\beta$\\ \hline
      $\gamma < \frac{5}{2}$  &  0  &  $\frac{1}{5- 2 \gamma}$\\
      $\frac{5}{2} < \gamma <\frac{7}{2}$  &  $\sqrt{\frac{\pi}{2m}}
                                             \frac{2\gamma-5}{4(\gamma-2)}$   &
                                                                               $\frac{1}{2
                                                                               \gamma-5}$\\
      $\frac{7}{2} < \gamma$  &  $\sqrt{\frac{\pi}{2m}}
                                             \frac{2\gamma-5}{4(\gamma-2)}$   &  $\frac{1}{2}$\\
    \end{tabular}
  \end{ruledtabular}
  \caption{Summary of critical properties at the HMF level of the modified CBV
    model on scale-free networks.}
\label{tab:critical_values}
\end{table}

\begin{figure}[t]
  \includegraphics{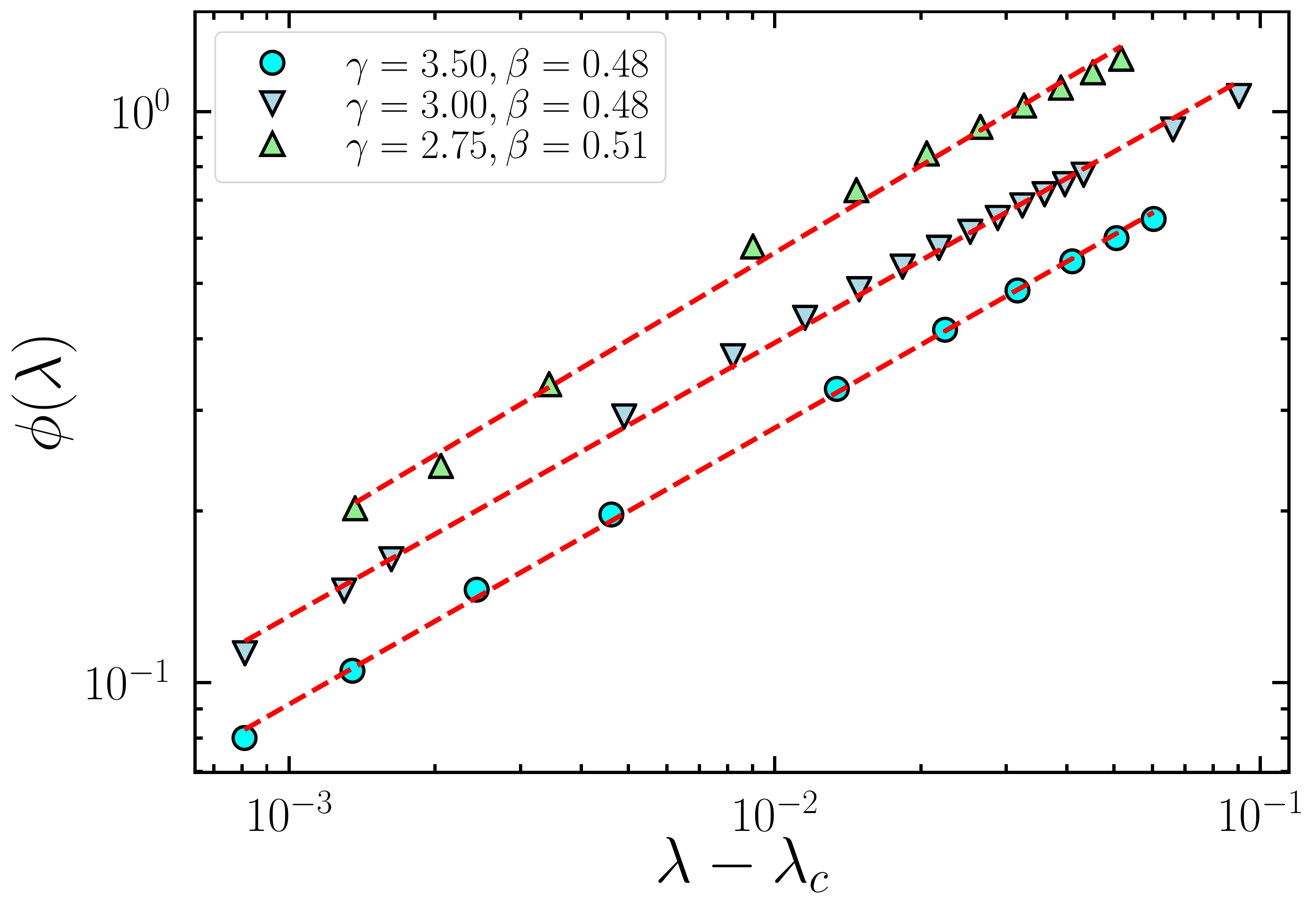}
  \caption{Plot of the order parameter as a function of the rescaled control
    parameter $\lambda$ in the vicinity of the tansition. The $\beta$ values
    quoted are obtained by means of a linear regression to the form
    $\phi(\lambda) \simeq (\lambda - \lambda_c)^\beta$, with $\lambda_c$ equal
    to the finite-size scaling extrapolation in the thermodynamic limit.  Plots
    have been shifted vertically for the sake of clarity.}
  \label{fig:beta_analysis}
\end{figure}

In any case, these $\beta$ exponents do not seem to capture the behavior of the
original CBV model on networks. In the case $\gamma < 5/2$, where the critical
point in the thermodynamic limits is $\lambda_c = 0$ (equivalently
$\eta_c \to \infty$), we should expect an order parameter decaying as a
power-law $\phi(\eta) \sim \lambda^\beta \sim \eta^{-\beta}$ for sufficiently
large $\eta$. This behavior is hampered by the fact that the effective critical
point $\eta_c(N)$ increases very slowly with $N$, see
Fig.~\ref{fig:peak_vs_N}. With the system sizes considered here, we cannot
observe the predicted behavior.  In the case $\gamma > 5/2$, the theory predicts
a finite threshold, so we can estimate the value of the $\beta$ exponent by
performing a linear regression to the form
$\phi(\eta) \sim (\lambda - \lambda_c)^\beta$, where $\lambda_c$ is obtained
from the finite-size scaling extrapolation to infinite network size. In the
analysis performed in Fig.~\ref{fig:beta_analysis} we obtain a very good fit to
this expression, with an exponent $\beta \simeq 1/2$, apparently independent of
the degree exponent $\gamma$.
This numerical value of the $\beta$ exponent is equal to the pure mean-field
prediction corresponding to $\gamma > 7/2$, in agreement with the result
obtained for $\gamma = 3.50$. For $\gamma = 3.00$ and $\gamma = 2.75$, our
numerical results are in strong disagreement with the HMF prediction, which in
this case would correspond to an exponent larger than $1$. The failure of HMF to
recover the numerical observation can be associated to different sources of
error. Firstly, the failure could also be attributed to the difference between
the numerically considered CBV model and the modified version actually
solved. The effect of a modulating $G(u)$ function that provides a binary output
could be enough to change the critical exponents of this flocking model.  Also,
it could be due to an intrinsic failure of HMF, as observed in other dynamical
processes, particularly epidemic spreading on scale-free
networks~\cite{Pastor-Satorras:2014aa}.

\section{Discussion}\label{sec:discussion}

In this paper we have contributed to the study of the effects of heterogeneous
social interactions in the behavior of flocking dynamics. To this end, we have
studied the dynamics of the flocking model proposed by Czir\'ok, Barab\'asi and
Vicsek (CBV), in which the interactions leading to the alignment of velocity are
mediated by a scale-free network, as observed in several animal social contact
networks. In opposition to the classical Vicsek model, in which velocity is a
vector, the CBV model considers a scalar velocity. Despite this difference, both
Vicsek and CBV models show an analogous behavior in power-law networks with
degree distribution $P(k) \sim k^{-\gamma}$: For mildly heterogeneous networks,
with $\gamma > 5/2$, the models show a true order-disorder transition, located
at a finite value of a noise intensity control parameter $\eta$. On the
contrary, for highly heterogeneous networks, with $\gamma < 5/2$, the transition
is absent in the thermodynamic limit, the systems being ordered for all physical
values of the noise intensity. Interestingly, despite the coincidence in the
position of the critical point, the Vicsek and CBV models show different
critical exponents ruling the growth of the susceptibility at the transition.

In order to gain insight into this phenomenology, we consider a modification of
the CBV model that can be solved within an heterogeneous mean-field (HFM)
approximation, typical for the analysis of dynamical processes on networks. In
our analysis, we confirm the results obtained numerically: The critical noise in
the modified CBV model scales as the degree moment $\av{k^{3/2}}$, and thus
experiences a transition between a finite and an infinite value, in the
thermodynamic limit, when $\gamma$ crosses the boundary $5/2$. The analytical
expression for the threshold is compared with numerical values, showing a
reasonably good fit. The HMF approximation allows also to compute the exponent
$\beta$ characterizing the ordered phase. The values obtained, however, are not
reproduced by numerical simulations. This failure of HFM can be attributed to
either insufficiently large network sizes, which do not allow to explore deep enough
into the critical phase, or to an intrinsic deficiency of the HMF approximation,
already observed in other dynamical processes.

Our results contribute to deepen the understanding of the dynamics of flocking
mediated by social interactions, with the introduction of a model amenable to a
simple analytical characterization, and that allows for more complex extensions,
such as the introduction of weights in the social ties. The analysis of these
kind of models can be improved, beyond the simplest HMF approximation, by
considering the full structure of the network in a quenched mean-field
approach~\cite{Castellano2010,Huang2017}. Finally, the observation of a
crossover behavior at a degree exponents $\gamma = 5/2$ might be conjectured to
be a general feature of flocking models based on velocity averaging. Further
investigation of this issue should be devoted to validate this
conjecture. Additionally, one can envisage potential applications in situations
where what really matters is not the particular direction of motion of a given
animal, but its dynamic response to particular stimuli that can be regarded as a
scalar quantity transmitted through a network of social contacts.  We can think
of situations, such as in the presence of predators, where assessing the
existence of movement might be more important than establishing an average
direction of motion. The results presented here would then have implications
regarding the ability of the system to respond collectively as a function of the
noise threshold.

\begin{acknowledgments}
  We acknowledge financial support from the Spanish MINECO, under projects
  FIS2016-76830-C2-1-P and FIS2016-76830-C2-2-P.  R. P.-S. acknowledges
  additional financial support from ICREA Academia, funded by the Generalitat de
  Catalunya.
\end{acknowledgments}

\end{document}